# Precise structure and polarization determination of $Hf_{0.5}Zr_{0.5}O_2$ with electron ptychography

**Authors:** Xiaoyue Gao[1,3†], Zhuohui Liu[2†], Bo Han[1,3], Xiaowen Zhang[1,3], Ruilin Mao[1,3], Ruochen Shi[1,3], Ruixue Zhu[1,3], Jiangbo Lu[4], Tao Wang[3], Chen Ge[2]* and Peng Gao[1,3]*.

**Affiliations:**

[1] International Center for Quantum Materials, School of Physics, Peking University, Beijing 100871, China

[2] Beijing National Laboratory for Condensed Matter Physics, Institute of Physics, Chinese Academy of Sciences, Beijing 100190, China

[3] Electron Microscopy Laboratory, School of Physics, Peking University, Beijing 100871, China

[4] School of Physics and Information Technology, Shaanxi Normal University, Xi'an 710119, China

†These authors contributed equally to this work.

*Corresponding author. Email: *gechen@iphy.ac.cn; pgao@pku.edu.cn*

## Abstract

$Hf_{0.5}Zr_{0.5}O_2$ (HZO) is a promising candidate for next generation ferroelectric memories and transistors. However, its ferroelectricity origin is still under debate due to the complex of its phase and microstructure in practical samples. In this study, we investigate the atomic structure of substrate-free HZO freestanding film with multislice electron ptychography, for which the ultra-high space resolution (up to ~25 pm) and capability to simultaneously image the cation and oxygen allow us to precisely determine the intrinsic atomic structures of different phases and reveal subtle changes among them. We clarify that the orthorhombic phase is ferroelectric with spontaneous polarization ~34±4 μC/cm² (corresponding to 56±6 pm in displacement) that is accurately measured through statistical analysis. Significant polarization suppression is observed near the grain boundary, while no distinguishable structural changes are detected near the 180° ferroelectric domain walls. Through the direct oxygen imaging of orthorhombic phase from the [111] zone axis, we quantify a substantial number of oxygen vacancies with a preferential distribution, which influences the polarization direction and strength. These findings provide fundamentals for HZO research, and thus lay a foundation for the design of high-performance ferroelectric devices.

## Introduction

Hafnium zirconium oxide ($Hf_{0.5}Zr_{0.5}O_2$, HZO) has gained widespread attention in recent years due to its robust ferroelectricity at room temperature and excellent compatibility with the complementary metal-oxide-semiconductor (CMOS) technology[1-3]. It is considered to be the most promising candidate for the next generation of ferroelectric random-access memories (FeRAMs) and ferroelectric field-effect transistors (FeFETs)[4-7]. For example, the enhanced and switchable ferroelectricity in one nm $Hf_{0.8}Zr_{0.2}O_2$ films grown on silicon, which is unattainable in the traditional perovskite oxides, offers new opportunities for polarization-driven ultrathin FeRAMs[4]. Additionally, the intriguing negative capacitance behavior of the ferroelectric HZO captured by integrating it into a heterostructure capacitor with a second dielectric layer, opens new avenues for reducing power consumption and improving switching speeds in FeFETs for future electronic devices[5].

However, the origin of ferroelectricity in HZO has been a subject of continuous debate[8-12]. The mainstream view is that the ferroelectricity in hafnium oxide films originates from the ferroelectric orthorhombic phase (OIII phase, space group *Pca2₁*) as proposed by Boscke[1]. With the help of scanning transmission electron microscopy (STEM), the existence of a non-centrosymmetric orthorhombic phase was later evidenced by the position averaged convergent beam electron diffractions (PACBED)[13, 14], annular bright-field (ABF)[15] and integrated differential phase-contrast (iDPC) images[11]. Besides the ferroelectric OIII phase, other mechanisms are also proposed[12, 16-20] since the OIII phase is metastable under room temperature and atmospheric pressure, and there exists a dozen phase structures for HZO. For instance, previous works also proposed that HZO's ferroelectricity originates from a polar rhombohedral phase[18-20]. Additionally, theoretical and experimental studies have shown that oxygen vacancy ($V_O$), which is an intrinsic defects in HZO, are strongly linked to ferroelectric properties[12, 16, 17, 21, 22], e.g., in theory a certain amount of $V_O$ can stabilize the ferroelectric orthorhombic phase[8, 21, 23], although atomic-scale experimental evidence is lacking.

Accurately analyzing the intrinsic atomic structures (cations and oxygen configurations) of HZO is crucial but challenging due to the polycrystalline nature and substrate effects in those previous studies[24-26]. So, traditional methods such as X-ray diffraction averages information from numerous grains with multi-phases[27]. Indeed, even within a single grain with typical size in nanometer, the surface, grain boundaries (GBs) and domain walls (DWs) would have different atomic-scale features, necessitating the high space resolution characterization such as electron microscopy (EM). However, for the most commonly used HAADF-STEM image, the oxygen is usually invisible in due to its low atomic number, while the common oxygen sensitive EM techniques such as ABF[28], iDPC[29] and negative spherical aberration imaging (NCSI)[30] are highly sensitive to imaging conditions such as specimen mistilt, thickness variation and residual aberrations, which can easily lead artifacts[31]. The recently developed multislice electron ptychography (MEP) offers superior space resolution (~20 pm) compared to HAADF and ABF (typically ~60 pm)[32-35], and can simultaneously image cation and oxygen positions, depth information and correct specimen mistilt[36, 37], being ideal for quantification of atomic structure. On the other hand, nanosized grains are often overlapped along the viewing direction and could add artifacts to images as the TEM foil specimen is typically tens of nanometers thick. Moreover, substrate effects such as lattice clamping[38] and interfacial charge screening[39] also prevent the intrinsic structure and polarization analysis. In contrast, recent advances in freestanding membranes can mitigate these issues and enable precise atomic characterization in HZO[11, 24, 40].

Here, we use MEP to achieve ultra-high-resolution imaging (~25 pm) of polycrystalline HZO freestanding films. We reveal the atomic structures of multiple phases and measure the intrinsic polarization magnitude of HZO as ~56 ± 6 pm, corresponding polarization is 34 ± 4 μC/cm$^2$. Significant polarization suppression is observed at the grain boundaries extends from the outermost to the fourth polar layers. The ferroelectric and antiferroelectric phases generally coexisted within one grain, while no distinguishable polarization changes is observed at the phase boundaries and 180° DWs. Direct oxygen imaging of orthorhombic phase along the [111] zone axis revealed preferentially distributed oxygen vacancies that likely enhance the polarization. These findings provide more basic insights into the ferroelectricity of fluorite-structured ferroelectrics and their properties.

## Results

### The atomic structures of HZO

The schematic of MEP technique is shown in Fig. 1a, which scans the sample with an overfocused probe and collected the diffraction patterns using a direct electron pixelated detector.

The 4D-STEM datasets were reconstructed using MEP algorithm[33-35, 41]. The 5-nm-thick HZO freestanding film (Fig. 1b) is studied here to eliminate the influence of external factors such as interface effects, grain overlap and sample preparation damage to ensure the precise structural analysis. The morphology distribution (Fig. 1c) of grains with average size ~10 nm is consistent with previous studies[24]. The reconstructed real-space resolution is ~25 pm (Fig. 1d), which is much better than that of HAADF image (~61 pm, in Supplementary Fig. 1b-d). The reconstructed probes as shown in Supplementary Fig. 2, also demonstrate the reliability of the results. Projected MEP images in Fig. 1e show the ferroelectric phase of intrinsic HZO is orthorhombic phase (space group $Pca2_1$, FE phase), the antiferroelectric (AFE) phase, monoclinic (M) phase (Supplementary Fig. 3 and 5) and monoclinic-like (M') phase, which are in excellent agreement with the simulated reconstructions, underscoring the accuracy and reliability of the imaging technique in capturing the complex phase structures within HZO grains. We observe antiferroelectric nature in the M phase along the [100] direction, with varying displacements between neighboring layers. Previously unreported {010}-Hf/Zr antiphase boundary (APB) defects in the M phase were also captured and the corresponding crystal structure is provided, showing a significant influence on the nearby polarization (details provided in Supplementary Fig. 3).

Based on the probed atomic structures with ultra-high spatial resolution (Supplementary Fig. 4), we further performed precise statistical analysis of atomic positions on these structures. The lattice parameters were analyzed based on the half lattice c, a1 and a2, as shown in Figure 2. As displacive ferroelectrics, the FE phase (Fig. 2a) has centrosymmetric non-polar oxygen atoms (O1) and non-centrosymmetric polar oxygen atoms (O2) with different atomic distances. The lattices of FE phase consistent with the statistics of reported experimental and computational data (Supplementary Table S1), i.e., the distance of O1 in the FE phase is 41±2 pm while O2 is 60±5 pm (Fig. 2b). We also discovered subtle differences between different phases, such as the slightly larger oxygen atomic distances in AFE phase (Fig. 2c and d) and M' phase (Fig. 2e and f) compared to the FE phase. These findings are also highly consistent with structural parameters provided in the literatures[42-46]. Compared to the parameters of O and M' phase, we found that there is ~ 2.7% tensile stress along *a*-axis at O2 positions in O phase leading the metal atoms to relax along *a*-axis and shear along the *c*-axis, resulting the transition to M' phase and the disappearance of polarization. Specifically, the angle between atomic layers of adjacent metals is ~75° in the M' phase (Fig. 2c), while 70° in the M phase, providing more details of the M'-M phase transformation including strain effects and atomic relaxations (discussed in Supplementary Fig. 5). The extracted structural parameters of cubic (C) and uncommon tetragonal (T) phase (indistinguishable from the *c*-axis zone) phases are consistent with previous studies (Fig. 2 g and h)[47].

**Polarization and localized phenomena near grain boundaries**

We imaged a ferroelectric grain from [010] axis zone (Fig. 3a) and mapped the distribution of displacements between cation and oxygen. Statistical analysis of the polarization displacements in O-phase grains (Fig. 3b) suggests that the polarization displacement is ~56 ± 6 pm with corresponding spontaneous polarization $P_s$ of 34±4 μC/cm$^2$ (details in Methods), consistent with the theoretical value[43]. The distribution near the zero point in the histogram represents statistics from the non-polarized layers, the value of in-plane displacement (0.8±3 pm) along *a*-axis indicates that the polarization direction is along *c*-axis (Supplementary Fig. 6b). Through studies of multiple grains, we observed that grains with uniform polarization in one direction (Fig. 3a) are rare in the pristine film. Most O-phase grains exhibit a mixture of FE and AFE phases (Fig. 3c), characterized by multiple 180° DWs. This is reasonable due to the close energy between the FE and AFE phases with the independently switchable characteristics of the polar layers[10]. The polarization magnitude mapping (Fig. 3d) of the grain shown in Fig. 3c shows that no detectable polarization variation at the 180° DWs (Supplementary Fig. 6), while the obvious polarization suppression near the grain boundary can be easily observed. Polarization suppression is not only observed at the grain boundaries in the [100] direction, but also exists in the [001] direction with a general polarization suppression phenomenon. Partial polarized oxygen ions near the boundaries are relaxed towards center position but still non-overlapping (Fig. 3c), while the lattice c shows no obvious suppression (Supplementary Fig. 6d) resulting in the polarization suppression in whole relaxed polar layer. From Fig. 3e and f, we find that the typically the polarization suppression occurs at the outermost three polar layers at the edge. The underlying mechanism of such polarization suppression is usually very complicated at a grain boundary. From the electron energy loss spectroscopy (EELS) measurement, we find that significant and widespread oxygen vacancies are present at the grain boundaries (Supplementary Fig. 7), i.e., the blurring of the doublets of the O-K edge in EELS indicates that higher concentration of $Vo$ located at GBs. The $V_O$ that alters the local charge states, should be closely with the polarization suppression near the boundary[48], although the full mechanism requires a more comprehensive study in future.

**The oxygen visualization of O-phase with [111] out-of-plane**

The [111] is the common dominant orientation in HZO thin films. Thus, imaging oxygen atoms along this direction is essential for understanding and optimizing the ferroelectric properties of HZO thin films. However, direct imaging of oxygen atoms from this direction challenging as the closest projected distance between oxygen and cation is less than 75 pm, which is typically too small for the conventional electron microscopy techniques[2, 24]. By using MEP, we imaged the oxygen atoms from the O-[111] orientation, as shown in Fig. 4a. The experimental image exhibits excellent consistency with the orthorhombic phase. Note that only the oxygen atoms of the polar layers are visible from the orientation and are displayed for clarify in the main text, while the full structure is shown in Supplementary Fig. 8.

As a common intrinsic defect in hafnium-based ferroelectric materials, oxygen vacancy $V_O$ play a crucial role in influencing the material's ferroelectric properties and device performance[21, 49]. Previous studies have shown that the O phase $HfO_2$ with a certain amount of oxygen vacancies exhibits enhanced stability and stronger polarization[21, 23, 50-52]. Consequently, we analyzed the projected image to quantify the presence of oxygen vacancy[53]. The presence of $V_O$ can be evident from the intensity differences between line-profiles in Fig. 4b. Notably, the intensity of the oxygen-deficient region, as indicated by the blue arrow, remains higher than that of the vacuum region. We conducted the corresponding intensity mapping of oxygen atoms (Fig. 4c) and find that the intensities are not uniform while the intensity of the metal atoms remains consistent (Supplementary Fig. 8c). From the quantitative analysis for the whole region in Supplementary Fig. 8, surprisingly, we find that significant differences in O occupation at different sites as shown in Fig. 4d. The phase distributions of the B and C polarized sites showed lower values compared to the A site, with the estimated $V_O$ concentration values of 0.15±0.15 and 0.27±0.14, respectively. The presence of $V_O$ at polar oxygen sites is expected to affect the ferroelectricity of HZO, i.e., compared to the $V_O$-free atomic structure (Fig. 4e), the presence of $V_O$ likely changes the polarization direction and enhance polarization magnitude (Fig. 4f). Therefore, the uneven or preferential distribution of $V_O$ can effectively modulate the ferroelectric properties of HZO. This may also help explain the past experiments that adjusting oxygen concentration and employing bombardment treatments can effectively enhance the performance of $HfO_2$-based ferroelectric films[16, 54].

## Discussion

Our high-resolution MEP images has precisely measured the atomic structures of HZO phases in freestanding films, and revealed the ferroelectric phase, domain wall structure, polarization suppression at grain boundaries, thus significantly enhancing our understanding of HZO properties especially for the ferroelectricity. With precise imaging including oxygen distances, we declare the origin of HZO intrinsic ferroelectric phase is orthorhombic phase (space group *Pca2₁*). Different from common perovskite ferroelectrics, this HZO ferroelectric phase consists of polar and non-polar layers. In polar layers, the polar displacements is measured to be 56±6 pm, corresponding ~34±4 µC/cm², which in excellent agreement with previous theoretical values[9, 10, 43]. However, most of the reported experimental values are below 30 $\mu C/cm^2$ [1, 55], e.g., it is 16 $\mu C/cm^2$ for the 5 nm HZO film in our study (Fig. 1c), which is likely due to coexistence of non-polar structures.

The 180° domain walls ubiquitously exist in the ferroelectric phase owing to the negligible energy difference between the FE phase and AFE phases. Across 180° domain walls, neither substantial changes in lattice constant nor polarization is observed, which is also different than the perovskite ferroelectrics. However, at boundaries, polarization is usually suppressed, e.g., up to ~ 13 pm (~23%) in displacement. The thickness of polarization suppression generally appears more than a single polar layer. These details are helpful for us to understand the screening mechanism in HZO devices and further optimize the sample preparation.

As the dominant growth orientation, the [111] zone axis of O phase, its polarization characteristics are crucial for the ferroelectric properties of HZO[2, 24, 56]. However, the tiny atomic distances from this orientation with closest distance below 75 pm, makes the traditional imaging challenging to determine the oxygen positions due to the limited space resolution. Our direct imaging of oxygen atoms with ultrahigh space resolution from the [111] zone axis has overcome this challenge and thus provided direct evidence for the presence of $V_O$. The capability to quantify oxygen vacancies can help us to precisely establish the relation between point defects and the ferroelectric properties. Our results are reminiscent of similar studies on ferroelectric polarization induced by ordered oxygen vacancies in the M phase[17, 22], as well as the use of oxygen concentration treatments to improve film performance[16, 23, 52].

## Conclusions

In summary, we investigate the intrinsic atomic structures and defects in HZO freestanding film with MEP. We experimentally determined the intrinsic polarization magnitude of HZO is ~56±6 pm. We find that the grain boundaries and oxygen vacancies can significantly influence polarization. Notably, substantial polarization suppression occurs around grain boundaries, extending across multiple polar layers. Furthermore, the direct imaging of oxygen atoms along the predominant [111] zone axis uncovered a preferential distribution of oxygen vacancies, which should enhance the polarization. These insights highlight the complexity of phase transitions and polarization phenomena in HZO, providing valuable insights into optimizing hafnium-based materials.

# Methods

**Sample Preparation.**

The HZO films with 5 nm thickness were synthesized as $HZO/La_{0.8}Sr_{0.2}MnO_3$ (LSMO) heterostructures on single crystal $SrTiO_3$ (STO) through pulsed laser deposition (PLD) method using a XeCl laser ($\lambda$= 308 nm) and the deposition temperatures was 780 ℃. The laser energy fluence was 1.75 $J/cm^2$, and the repetition rates used for both films were the same (2 Hz). The deposition rate of films was calibrated by X-ray Reflection. After growth, the HZO films were released from the STO substrate by selective etching of the sacrificial LSMO layer. The unclamped thin films were removed from the substrate by slowly dipping it into deionized water. A homemade copper wire loop was used to catch the freestanding HZO membranes from the water surface, held in place by a thin water layer inside the loop. The films were then transferred to target substrates on a 95 ℃ hot plate. The freestanding HZO films remained on the TEM grids after water evaporated.

**Experimental 4D-STEM datasets acquisition.**

The 4D-STEM experiments were operated at 300 kV on a double-aberration-corrected JEOL JEM-ARM300F2 microscope. Considering the thickness of film, convergence semi-angle ($\alpha$) of 32.8 mrad was selected for all datasets[57]. We adopted an overfocus probe ~ 10 nm above the sample surface to collect the datasets. The 4D-STEM datasets were recorded with 0.819 mrad/pixel by using a Medipix3RX MerlinEM direct electron pixelated detector with 256×256 pixels. The datasets were acquired using a dynamic range of 12-bit in electron-counting mode with 0.617 ms dwell time per diffraction pattern with scan step size about 0.304 Å. The first two rows of raw datasets were deleted to avoid scan noise. The HAADF images were recorded on JEOL JEM-ARM300F2 microscope operated at 300 kV.

**Ptychographic reconstructions and data analysis.**

Ptychographic reconstructions were performed using the mixed-state multislice algorithm available in the fold-slice package[33, 34], using 5 slices with slice thickness of 1 nm. For each dataset, multiple reconstructions were iterated to improve the resolution with each undergoing more than 600 iterations. The moderate regularization parameters ranging from 1 to 0.5 were applied for high-quality reconstructions [35]. Based on the complex structures of HZO, AtomAI[58] and Atomap[59] were used to fit and analyze the atomic positions, intensity and structures. The atom positions were predicted using a custom-trained model in AtomAI and then Atomap was used for 2D-Gaussian fitting to refine the positions ensuring precision. The O1 distances were accurately determined by fitting column with Gaussian functions. For polarization analysis, the obtained relative displacement $\delta$ of cations and anions was used to calculate polarization by $P_s = \frac{e}{V}\Sigma\delta_i Z_i$, where $e$ is the electronic charge, $V$ is the volume of unit cell and $Z$ is born charge values[60]. The Born charge value for the A site is 5.15, which is derived as the mean value of 5.0 for Hf and 5.3 for Zr[61, 62]. We choose the smallest polarization in the outermost polar layer to obtain the suppression maximum of 13 pm (indicated by blue arrow in Fig. 4d).

**Simulation of 4D-STEM datasets.**

All simulated 4D-STEM datasets were implemented with abTEM [63] using Kirkland's parametrization of atomic potentials. Thirty frozen phonon configurations were averaged to account for thermal diffuse scattering. Parameters such as beam energy, scan step, defocus and convergence semi-angle from the experiments were used in the simulations.

**STEM-EELS.**

The energy loss spectra were acquired at a Nion U-HERMES200 electron microscope equipped with both the monochromator and the aberration corrector operated at 60 kV. The probe convergence semi-angle was 35 mrad, while the collection semi-angle was 25 mrad. The core-level EEL spectrum was recorded as spectrum image with 1.6 nm × 8 nm and 0.08 nm per pixel. EEL spectra were processed using the Gatan Microscopy Suite and custom-written MATLAB code. The spectra were summed along the direction parallel to the interface to obtain the line-scan data with a good signal-to-noise ratio.

**P-E hysteresis measurement.**

The Pt/LSMO/HZO capacitors were measured using a precision multiferroic analyzer (RADIANT Tec. Inc.). The frequency of triangular pulses used to switch the polarization of ferroelectric HZO layer is 10 kHz. P–E measurement is performed using tungsten probe.

## Acknowledgments

We acknowledge Electron Microscopy Laboratory of Peking University for the use of electron microscopes and the High-performance Computing Platform of Peking University for providing computational resources. We acknowledge the assistance of Prof. Zhen Chen from the School of Physical Sciences, University of Chinese Academy of Sciences, Beijing, China; We thank the discussion of Yi Jiang from the Advanced Photon Source, Argonne National Laboratory, USA.

**Funding:** This work was supported by the National Natural Science Foundation of China (52125307 to P.G., 12222414 to C.G.) and the open research fund of Song-shan Lake Materials Laboratory (2022SLABFK03). P.G. acknowledges the support from the New Cornerstone Science Foundation through the XPLORER PRIZE.

**Author contributions:** X.Y.G., Z.H.L. contributed equally to this work. P.G. conceived the project; Z.H.L. grew the sample with the guidance of C.G..; X.Y.G. performed the ptychographic experiment, simulation, reconstruction and data analyses with the assistance of B.H., R.C.S., R.X.Z., R.L.M., J.B.L and T.W.; X.W.Z. assisted the STEM-EELS experiment. X.Y.G. wrote the manuscript under the direction of P.G.; All the authors contributed to this work through useful discussion and/or comments to the manuscript.

**Data and materials availability:** All data needed to evaluate the conclusions in the paper are available in the main text or the supplementary materials. Additional data related to this paper may be requested form the authors.

**Competing interests:** The authors declare no competing interests.

## Supplementary Information

Supplementary Information is available for this paper.

Figs. S1 to S8

Table S1

References

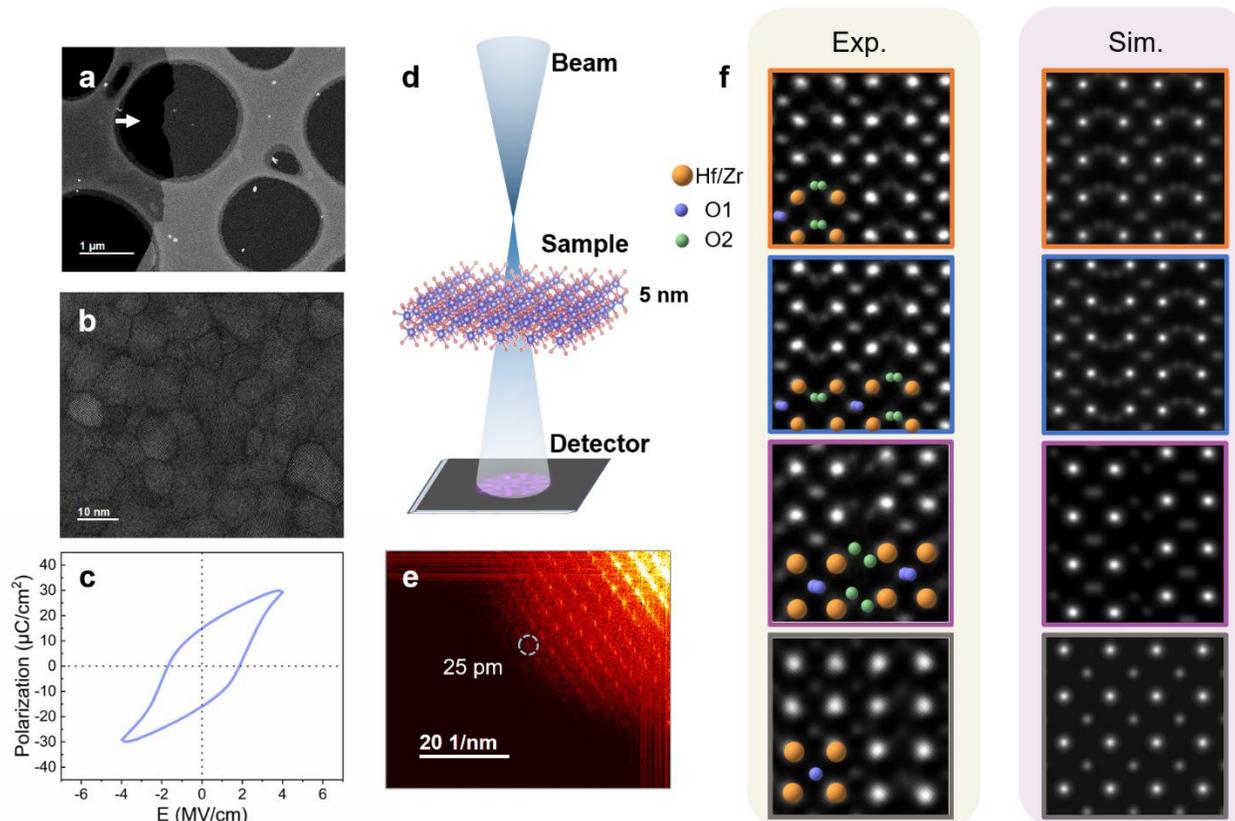

**Fig. 1. Imaging atomic structures of HZO freestanding thin film. a,** Low-magnification plan-view HAADF-STEM of HZO freestanding film with thickness ~ 5 nm. The white arrow indicates the edge sample (on the right-side of the image). **b,** The HAADF-STEM shows grain distribution. **c,** Polarization–electric field (P–E) hysteresis loops at 10 kHz. **d,** Schematic of multislice electron ptychography. **e,** Fourier transform of the projected phase image in Supplementary Fig. 1e. The dashed circle denotes a resolution 25 pm in real space. **f,** Projected phase images from multislice electron ptychographic reconstructions of experimental (left side) and simulated (right side) datasets, overlapped with atomistic structure models. From top to bottom are the FE phase, AFE phase, M' phase, and C/T phase, respectively.

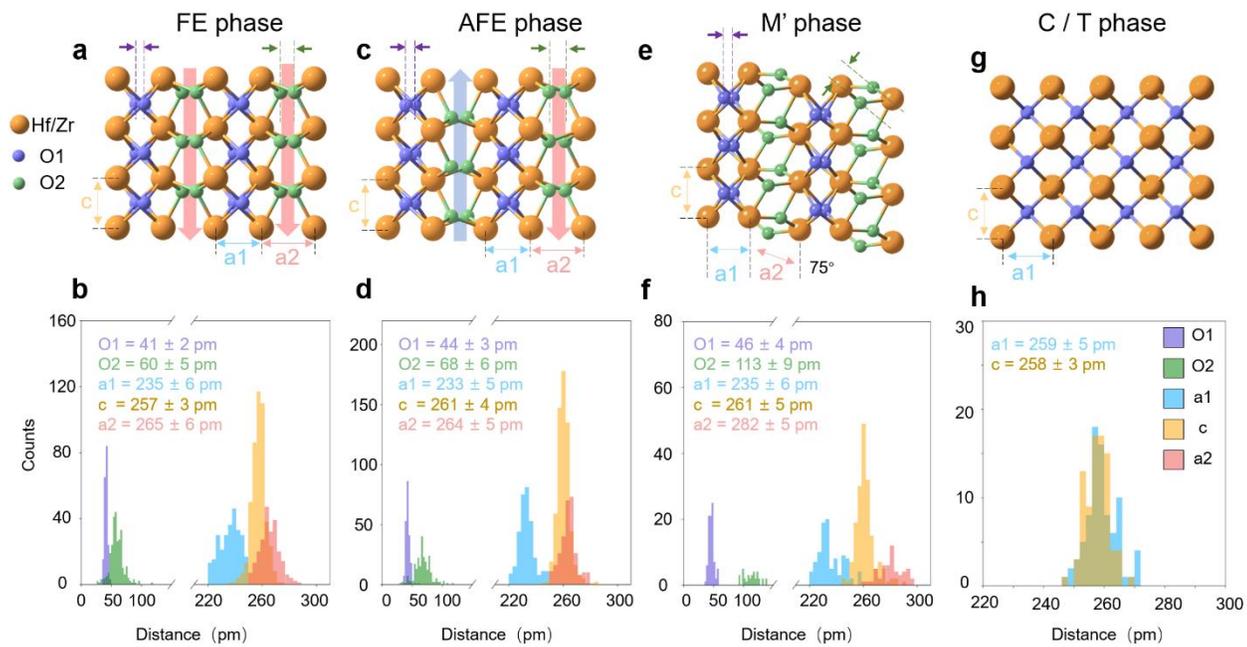

**Fig. 2. Determination of atomic structure and lattice parameters from reconstruction.** Atomic structures of the O-FE phase (**a**), O-AFE phase (**c**), M' phase (**e**) and C or T phases (**g**) based on experimental reconstructions. The red and blue arrows represent downward and upward polarization directions respectively. Histograms of lattice parameters for each phase shown as (**b**) for (**a**), with (**d**) for (**c**), (**f**) for (**e**) and (**h**) for (**g**). The half lattice c, a1 and a2 are labeled with colors.

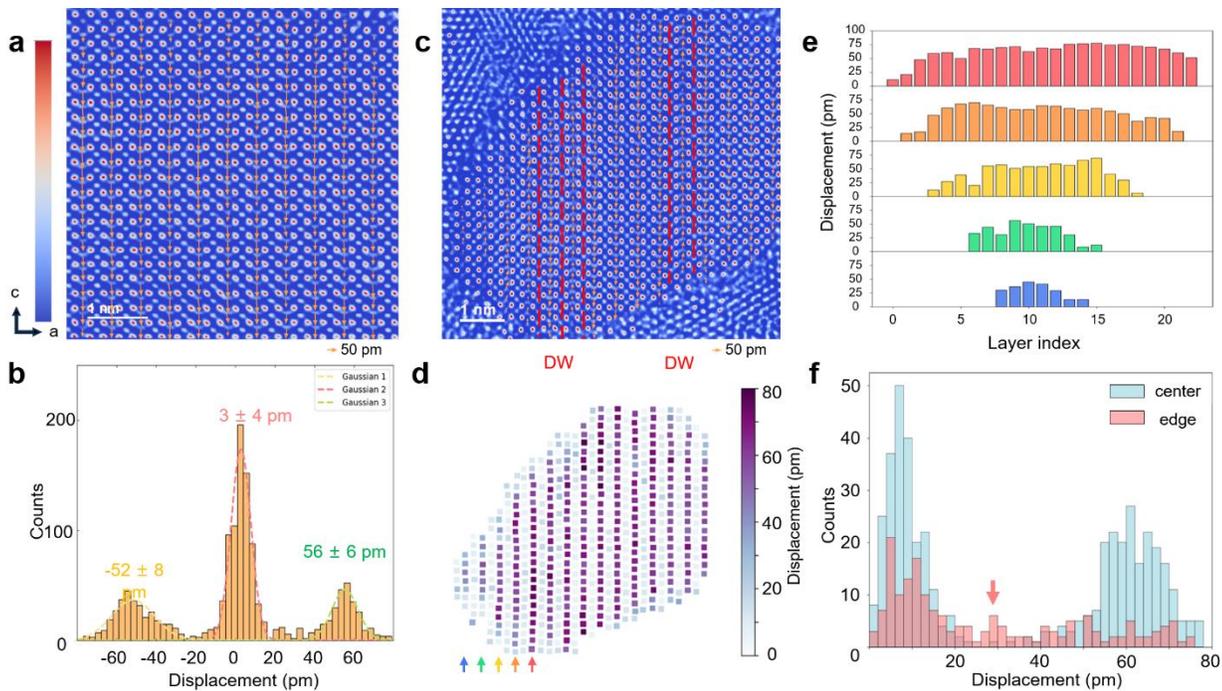

**Fig. 3. Ferroelectric polarization mapping. a,** Projected MEP phase image of a FE grain along [010] axis zone with the displacement arrows overlapped. **b,** Polarization displacements in O-phase HZO grains. The fit curves represent Gaussian fitting. **c,** Projected MEP phase image from a FE-AFE mixed grain with the displacement arrows overlapped. **d,** The corresponding displacement magnitude map of (**c**). **e,** The polarization displacements of from different atomic layer labeled by the colored arrows in (**d**). **f,** The displacement distributions from the edge (red, three outermost layers) and center (blue, the rest part) of grain shown in (**c**).

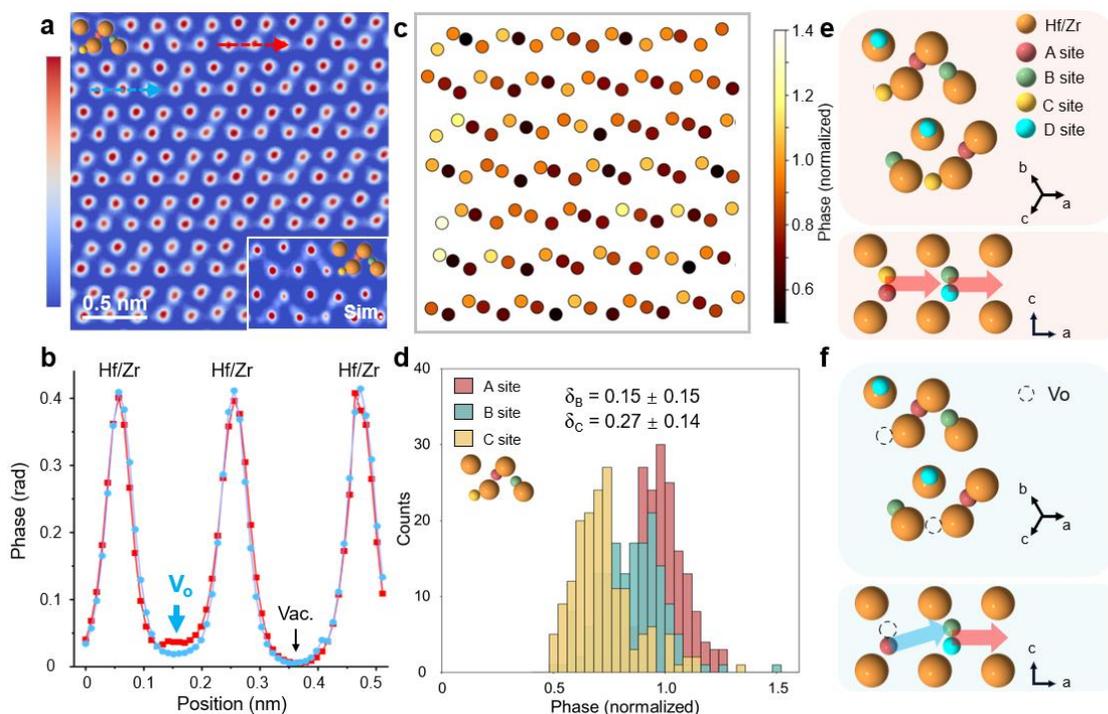

**Fig. 4. Oxygen Visualization of along the [111] axis. a,** Projected MEP images of experiment and simulation (inset). **b,** Intensity linecuts along the corresponding color dashed arrows marked in (**a**). **c,** The colored phase mapping of oxygen columns in (**a**). **d,** Corresponding phase histogram from the oxygen sites along [111] axis. The full image is shown in Supplementary Fig. 8e, Schematic of FE phase without centrosymmetric nonpolar O1 for clarity. The complete atomic model along [111] axis zone is shown in Supplementary Fig. 8. **f,** Schematic of FE phase with $V_O$. The color arrows shown in (**e**) and (**f**) indicate polarization direction. In (**f**), the magnitude of blue arrow should be larger than that of the red one.

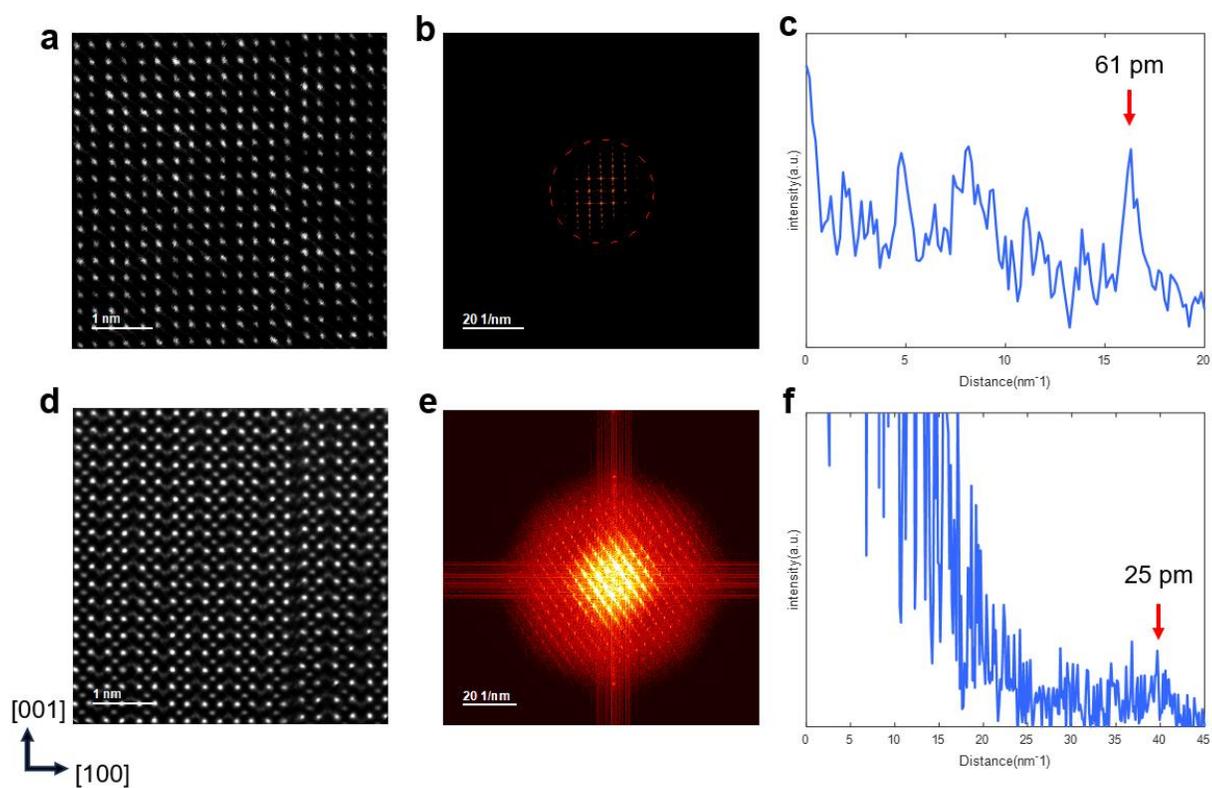

**Fig. S1. Sample morphology and comparison of experimental HAADF-STEM imaging and MEP imaging. a,** HAADF image with corresponding Fourier transform (**b**), the linecut (**c**) from the spectrum (**b**). **d,** Projected MEP image with corresponding Fourier transform (**e**), the linecut (**f**) from the spectrum (**e**).

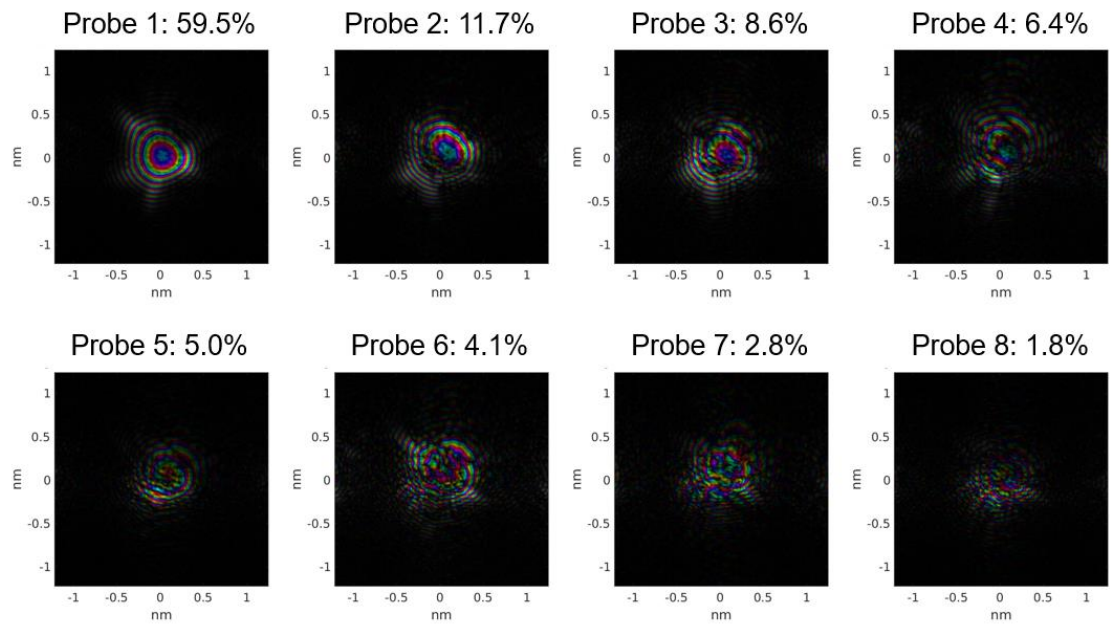

**Fig. S2. Visualization of reconstructed Probes. The brightness represents the magnitude of the probes while the color represents the phase. The ratios of every probe state shown as percentage.**

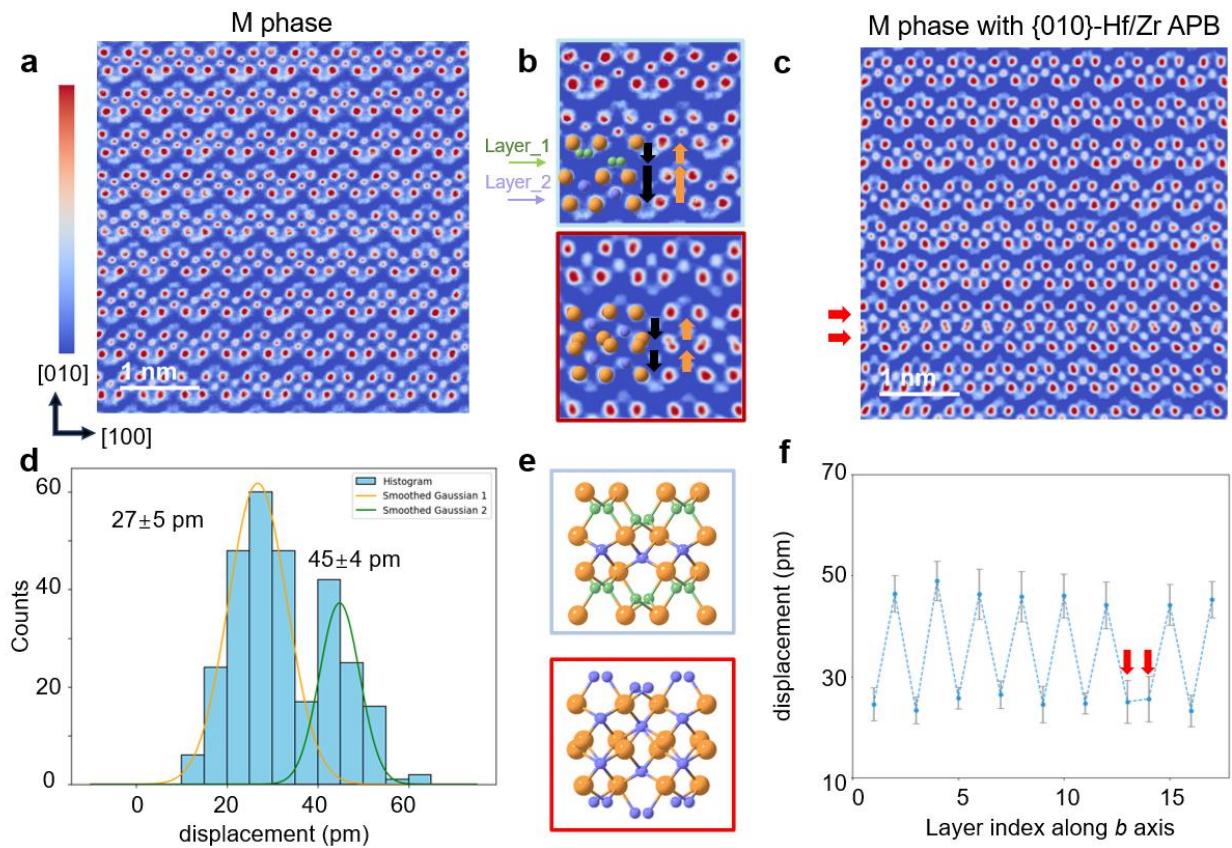

**Fig. S3. M phase and defects along [001] axis zone with corresponding polarization histograms. a,** Projected phase image of the M phase along the [001] axis zone, revealing antiferroelectricity along the [100] direction (top part in **b**), with neighboring layers displaying different displacements of 27±5 pm and 45±4 pm (**d**). **c,** Unreported {010}-Hf/Zr APB defects (indicated by red arrows) in the M phase, where the metal atoms in the middle tier slide, leading to a shift of the O2 atoms in Layer_1 closer to the O1 atoms. This shift reduces the polarization in Layer_1 to the same level as in Layer_2 (bottom part in **b**). **e,** The atomic models of M phase (top part) and the APB defects (bottom). **f,** The polarization line profile from (**c**) along *b* axis.

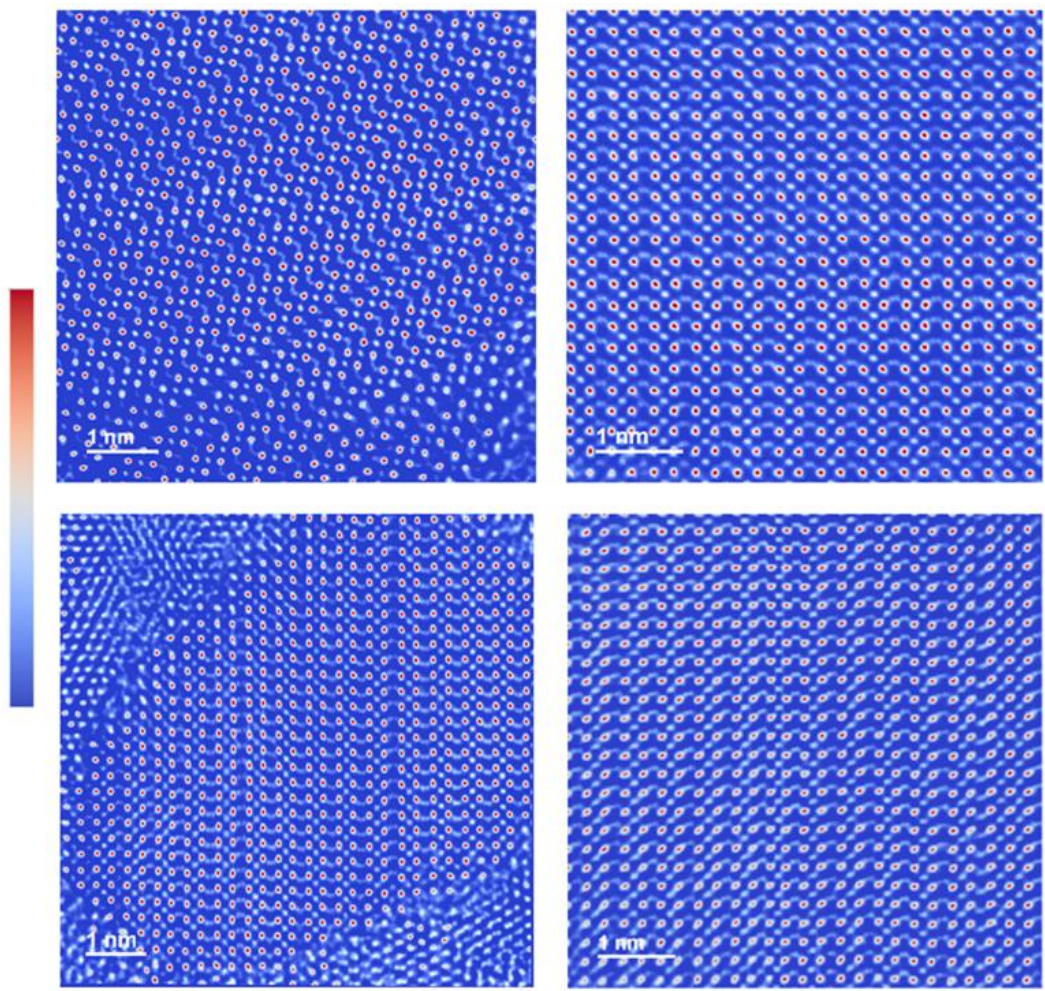

148    **Fig. S4. Reconstructed images for structure analysis.**

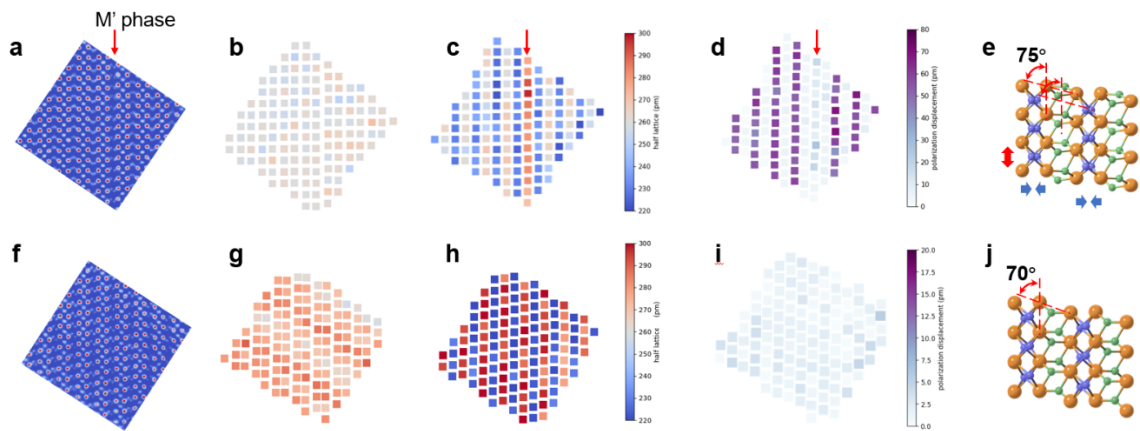

**Fig. S5. M' and M phase along [010] axis zone with corresponding parameter mappings and atomic models. a,** M' phase in O-phase grain with corresponding half c (**b**), a (**c**) and polarization (**d**) mapping. **e,** The M' phase atomic model based on (**a**). **f,** Bulk M phase corresponding c (**g**), a (**h**) and polarization (**i**) mapping. **j,** M phase atomic model based on (**f**). The (half) parameters of M phase are: c is $275 \pm 7$ pm, a1 is $221 \pm 5$ pm and a2 is $293 \pm 7$ pm. The angle between the atomic layers of adjacent metals in the M' phase is 75°, while 70° in the M phase. As the metal atoms (Hf/Zr) in the M' phase slip towards the M phase grain, compressive strain in the *c*-axis direction is released. Meanwhile, the metal atoms in the unit cell further relaxed towards the O1 atoms along the *a*-axis direction, as indicated by the blue arrows, resulting negligible strain in the a-axis direction but compressive strain within sub cell of a1 (**h**). This shift distorts the horizontal O1 atoms (**e** and **j**).

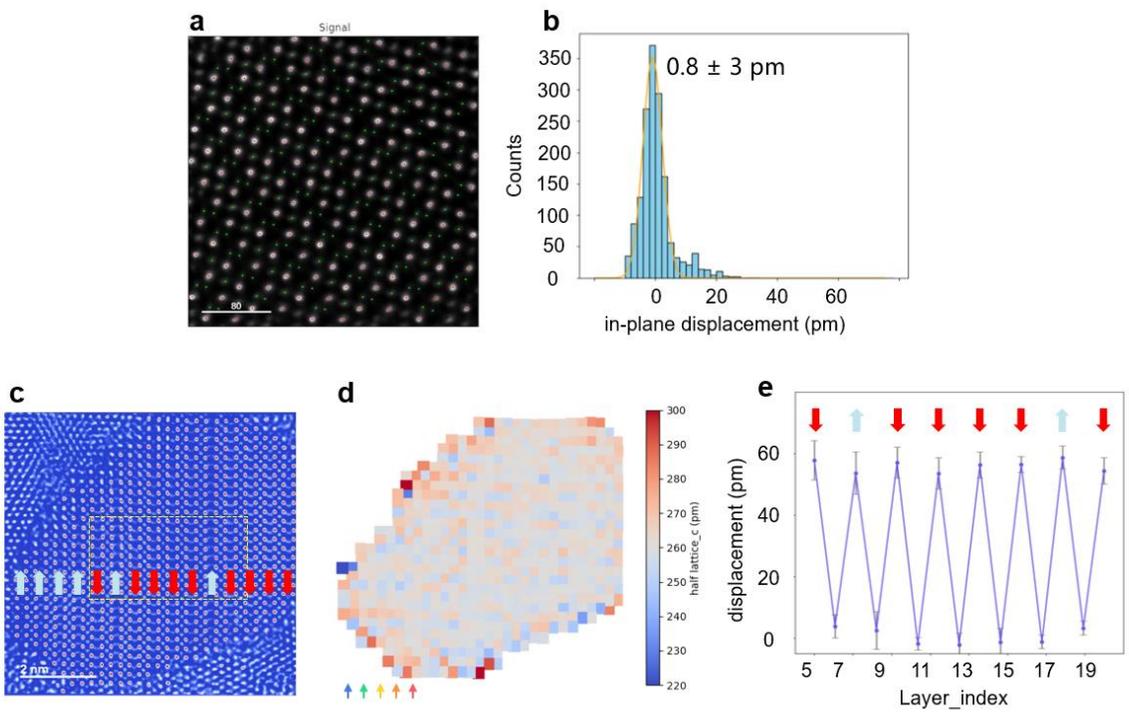

**Fig. S6 . Analysis of Reconstructed images. a,** represented fitting result of atomic positions. **b,** The histogram of in-plane polarization displacement. **c,** The image of grain with 180° DWs, as the colored arrows indicated the directions of polarization. **d,** The half lattice c map of (**c**). The arrows indicate the same region in main text Fig. 3d. **e,** The polarization line chart from yellow dashed-line region from (**c**) with error bars.

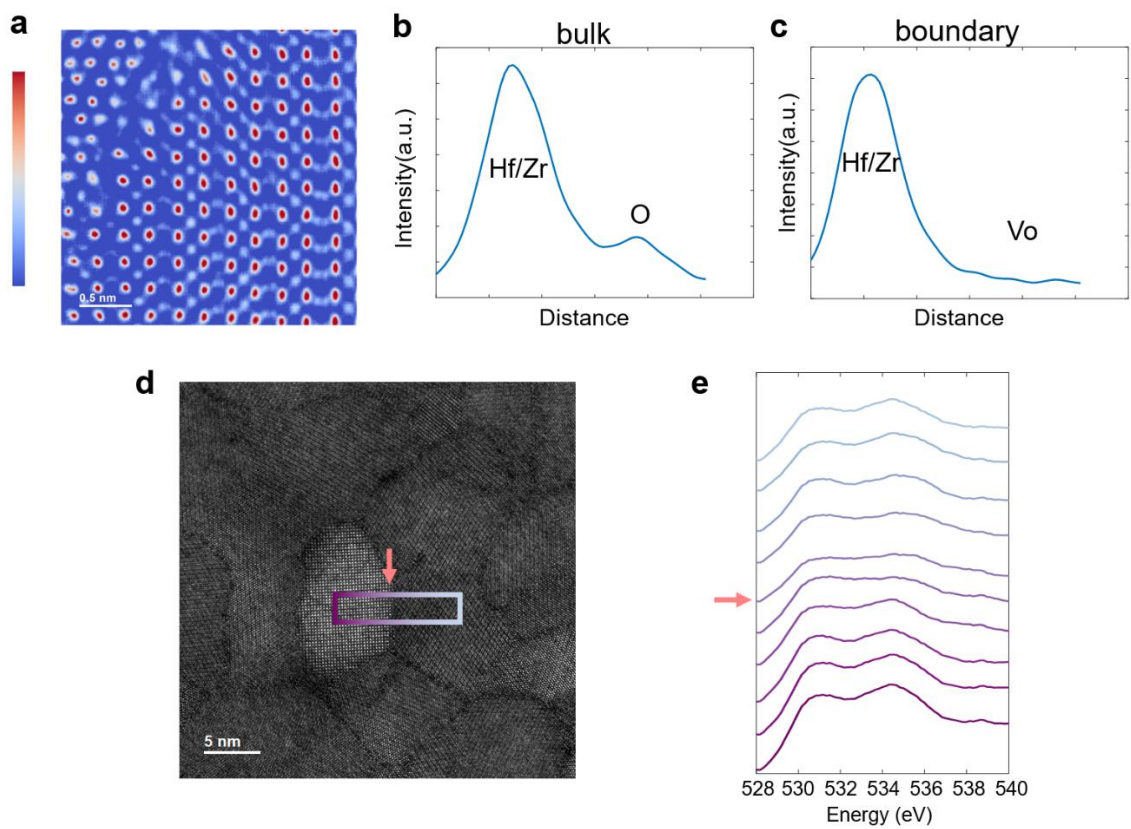

**Fig. S7. $V_O$ near grain boundary. a,** Enlarged image near the grain boundary from Fig. 3c. Bulk (**b**) and boundary (**c**) intensity profile show the exist of $V_O$. **d,** The HAADF image with corresponding region of eels. **e,** O-K EELS spectrum. Peak shape changes indicated by the red arrow also suggest the higher concentration of $V_O$.

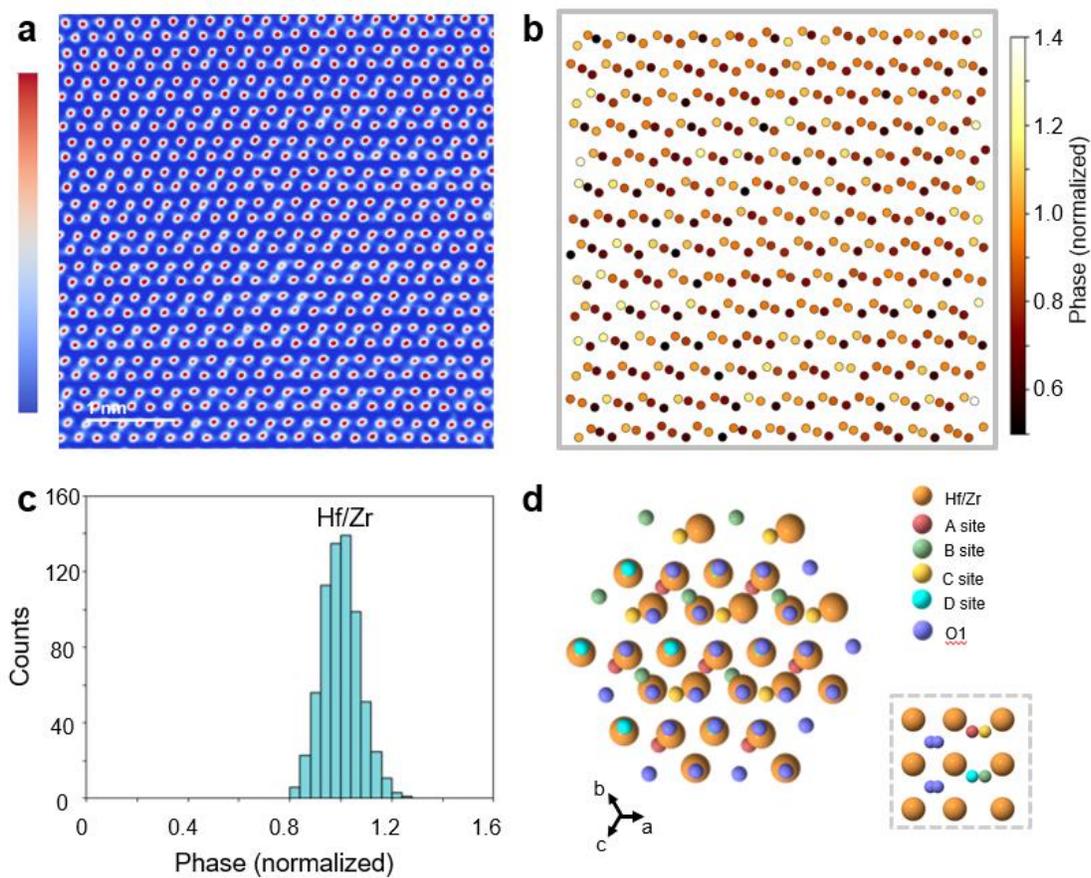

**Fig. S8. Analysis and structure along the [111] axis. a-b**, The full projected MEP image and intensity mapping of Fig. 4**a** and **b**. **c,** The phase intensity histogram of metal atoms. **d,** The completed atomic model of FE phase.

172 **Table S1**

173 **Lattices statistics of O phase structure**

174 (*Exp.x* from experimental references, *Com.x* from computational references)

| Ref. | lattice a (Å) | lattice c (Å) | Ref. | lattice a (Å) | lattice c (Å) |
|---|---|---|---|---|---|
| *Exp.1* | 5.07 | 5.08 | *Com.1* | 5.06 | 5.14 |
| *Exp.2* | 5.07 | 5.08 | *Com.2* | 5.08 | 5.11 |
| *Exp.3* | 5.07 | 5.08 | *Com.2* | 5.08 | 5.11 |
| *Exp.4* | 5.07 | 5.08 | *Com.3* | 4.92 | 4.96 |
| *Exp.5* | 5.01 | 5.05 | *Com.3* | 5.09 | 5.14 |
| *Exp.6* | 5.06 | 5.07 | *Com.4* | 5.16 | 5.18 |
| *Exp.7* | 5.06 | 5.07 | *Com.5* | 5.06 | 5.08 |
| *Exp.8* | 5.10 | 5.00 | *Com.5* | 5.13 | 5.16 |
| *Exp.9* | 5.10 | 5.00 | *Com.6* | 5.05 | 5.08 |
| *Exp.10* | 5.02 | 5.04 | *Com.7* | 5.08 | 5.10 |
| *Exp.11* | 5.10 | | *Com.7* | 5.04 | 5.07 |
| *Exp.12* | 5.04 | 5.09 | *Com.7* | 5.12 | 5.14 |
| *Exp.13* | 5.01 | 5.06 | *Com.8* | 5.05 | 5.07 |
| *Exp.13* | 5.02 | 5.07 | *Com.9* | 5.02 | 5.04 |
| *Exp.14* | 5.10 | | *Com.10* | 5.01 | 5.08 |
| *Exp.16* | 5.00 | 5.09 | *Com.11* | 5.04 | 5.07 |
| *Com.1* | 5.07 | 5.20 | *Com.11* | 5.13 | 5.16 |
| *Com.1* | 5.07 | 5.15 | *Com.12* | 5.05 | 5.08 |

175